\documentclass[12pt]{article}
\usepackage{graphicx}
\usepackage{epsfig}
\usepackage{amsmath,amssymb,hyperref,enumerate}
\usepackage{pstricks}
\usepackage{slashed}
\usepackage{amsfonts}
\usepackage{url}
\usepackage{color}
\usepackage{appendix}
\usepackage{bm}
\usepackage{subfig}
\usepackage{cite}
\hypersetup{colorlinks,citecolor= black,linkcolor= black}
\definecolor{nicered}{rgb}{0.7,0.1,0.1}
\definecolor{nicegreen}{rgb}{0.1,0.5,0.1}

\usepackage{setspace}

\allowdisplaybreaks

\def\({\left(}
\def\){\right)}
\def\[{\left[}
\def\]{\right]}

\begin{document}

\thispagestyle{empty}

\onehalfspacing

\begin{flushright}
NSF-KITP-14-175\\
OSU-HEP-14-11\\
\end{flushright}

\renewcommand{\thefootnote}{\fnsymbol{footnote}}

\begin{center}
\LARGE {Warm Dark Matter in Two Higgs Doublet Models
}
\end{center}

\vspace{0.5cm}

\begin{center}

{\large  K.S. Babu$^{a,b}$\footnote{Email:
babu@okstate.edu}, Shreyashi Chakdar$^{a,b}$\footnote{Email: chakdar@okstate.edu} and Rabindra N. Mohapatra$^c$\footnote{Email: rmohapat@umd.edu}}

\end{center}

\begin{center}

\it{ $^a$Kavli Institute for Theoretical Physics, University of California,\\ Santa Barbara, CA 93106, USA}
\vspace*{0.1in}

\it{ $^b$Department of Physics, Oklahoma State University,\\
Stillwater, Oklahoma 74078, USA}
\vspace*{0.1in}

\it {$^c$Maryland Center for Fundamental Physics and Department of Physics, \\University of Maryland, College Park, MD 20742, USA
}

\end{center}

\renewcommand{\thefootnote}{\arabic{footnote}}
\setcounter{footnote}{0}

\bigskip

\begin{abstract}

We show that a neutral scalar field, $\sigma$, of two Higgs doublet extensions of the Standard Model incorporating the
seesaw mechanism for neutrino masses can be identified as a consistent {\it warm} dark matter candidate with a mass of order keV. The relic density of $\sigma$ is correctly reproduced by virtue of the late decay of a right-handed neutrino $N$ participating in the seesaw mechanism.  Constraints from cosmology determine the mass and lifetime of $N$ to be $M_N \approx 25 ~{\rm GeV} - 20 ~{\rm TeV}$ and $\tau_N \approx (10^{-4} - 1)$ sec.  These models can also explain the 3.5 keV $X$-ray anomaly in the extra-galactic spectrum that has been recently reported in terms of the decay $\sigma \rightarrow \gamma \gamma$. Future tests of these models at colliders and in astrophysical settings are outlined.

\end{abstract}

\newpage

\section{Introduction}

One of the simplest extensions of the Standard Model is the addition of a second Higgs doublet to its spectrum.  A second Higgs doublet appears naturally in a variety of well motivated scenarios that go beyond the Standard Model.  These include supersymmetric models \cite{susy},  left-right symmetric models \cite{lr}, axion models \cite{dfsz} and models of spontaneous CP violation \cite{lee}, to name a few. These models have the potential for rich phenomenology that may be subject to tests at colliders and in low energy experiments.  A notable feature of these models is the presence of additional scalar states, two neutral and one charged, which may be accessible experimentally at the LHC.  Naturally, two Higgs doublet models have been extensively studied in the literature \cite{review}.

In this paper we focus on certain cosmological and astrophysical aspects of the two Higgs doublet models in a regime that has not been previously considered. It is well known that no particle in the Standard Model can fit the observed properties of the dark matter in the universe inferred from astrophysical and cosmological data.  New particles are postulated to fulfill this role.  Two Higgs doublet models do contain a candidate for dark matter in one of its neutral scalar bosons.  It is generally assumed that this particle, which is stable on cosmological time scales owing to an approximate (or exact) symmetry, is a {\it cold} dark matter candidate with masses in the several 100  GeV range\cite{ma,inert}.  These particles annihilates into lighter Standard Model particles in the early thermal history of the universe with cross sections of order picobarn.  In this paper we  show that there is an alternative possibility where the extra neutral scalar boson of these models can have mass of the order of a keV and be identified as a {\it warm} dark matter candidate. This scenario is completely consistent with known observations and would have distinct signatures at colliders as well as in cosmology and astrophysics, which we outline here.

The $\Lambda$CDM cosmological paradigm, which assumes a significant cold dark matter component along with a dark energy component in the energy density of the universe, has been immensely successful in confronting cosmological and astrophysical data over a wide range of distance scales, of order Gpc to about 10 Mpc.  However, at distance scales below a Mpc, cold dark matter, which has negligible free--streaming velocity, appears to show some inconsistencies. There is a shortage in the number of galactic satellites observed compared to CDM $N$--body simulations;
density profiles of galactic dark matter haloes are too centrally concentrated in simulations
compared to data; and the central density profile of dwarf galaxies are observed to be shallower than predicted by CDM \cite{CDM}. These problems can be remedied if the dark matter is {\it warm}~\cite{WDM}, rather than cold.  Warm dark matter (WDM) has non-negligible free--streaming velocity, and is able to wipe out structures at distance scales below a Mpc, while behaving very much like CDM at larger distance scales.  This would alleviate the small scale problems of CDM, while preserving its
success at larger distance scales. The free streaming length of warm dark matter can be written down very roughly as \cite{kusenko}
\begin{equation}
R_{fs} \approx 1 \,{\rm Mpc} \left({{\rm keV} \over m_\sigma}\right) \left({\langle p_\sigma\rangle \over 3.15 T}\right)_{T \approx {\rm keV}}~,
\end{equation}
where $m_\sigma$ is the dark matter mass and $\langle p_\sigma \rangle$ its average momentum.  For a fully thermalized
WDM, $\langle p_\sigma \rangle = 3.15 T$.  In the WDM of two Higgs doublet model, as we shall see later, $\langle
p_\sigma \rangle/(3.15 T) \simeq 0.18$, so that an effective thermal mass of $\sigma$, about six times larger than
$m_\sigma$ can be defined corresponding to fully thermalizd momentum distribution.  For $m_\sigma$ of order few keV, we see that the free--streaming length is of order
Mpc, as required for solving the CDM small scale problems.  Note that structures at larger scales would not be
significantly effected, and thus WDM scenario would preserve the success of CDM at large scales.

The WDM candidate of two Higgs doublet extensions of the Standard Model is a neutral scalar, $\sigma$,  which can have a mass
of order keV.  Such a particle, which remains in thermal equilibrium in the early universe
down to temperatures of order 150 MeV through weak interaction
processes (see below), would contribute too much to the energy density of the universe, by about a factor of 34
(for $m_\sigma = 1$ keV).  This unpleasant situation is remedied by the late decay of a particle that dumps entropy into other species and heats up the photons relative to $\sigma$.  A natural candidate for such a late decay is a right-handed neutrino $N$ that takes part in neutrino mass generation via the seesaw mechanism.  We find that for $M_N = (25 ~{\rm GeV}- 20~{\rm TeV})$,
and $\tau_N = (10^{-4} -1)$ sec. for the mass and lifetime of $N$, consistency with dark matter abundance can be realized.  Novel signals for collider experiments as well as for cosmology and astrophysics for this scenario are outlined.  In particular, by introducing a tiny breaking of a $Z_2$ symmetry that acts on the second Higgs doublet and makes the dark matter stable, the decay $\sigma \rightarrow \gamma \gamma$ can occur with a lifetime longer than the age of the universe.
This can explain the recently reported anomaly in the $X$-ray spectrum from extra-galactic sources, if $m_\sigma = 7.1$ keV
is adopted, which is compatible with other WDM requirements. This feature is somewhat analogous to the proposal of Ref. \cite{bm} where a SM singlet scalar which coupled
very feably with the SM sector played the role of the 7.1 keV particle decaying into two photons. The present model with $\sigma$ belonging to a Higgs doublet has an entirely different cosmological history; in particular $\sigma$ interacts with the weak gauge bosons with a coupling strength of $g^2 \sim {\cal O} (1)$ and remains in thermal equilibrium in the early universe
down to $T \approx 150$ MeV, while the singlet scalar of Ref. \cite{bm} was never thermalized.

The rest of the paper is organized as follows.  In Sec. 2 we describe the two Higgs doublet model for warm dark matter.
Here we also study the experimental constraints on the model parameter.  In Sec. 3 we derive the freeze-out temeprature of
the WDM particle $\sigma$ and compute its relic abundance including the late decays of $N$.  Here we show the
full consistency of the framework.  In Sec. 4 we analyze some other implications of the model.  These include supernova energy loss, dark matter self interactions, 7.1 keV $X$-ray anomaly, and collider signals of the model.  Finally in Sec. 5 we conclude.

\section{Two Higgs Doublet Model for Warm Dark Matter}

The model we study is a specific realization of two Higgs doublet models that have been widely studied in the context
of dark matter \cite{ma,inert}.  The two Higgs doublet fields are denoted as $\phi_1$ and $\phi_2$.  A discrete
$Z_2$ symmetry acts on $\phi_2$ and not on any other field.  This $Z_2$ prevents any Yukawa couplings of $\phi_2$.
While $\phi_1$ acquires a vacuum expectation value $v \simeq 174$ GeV, $\langle\phi_2^0\rangle = 0$, so that the $Z_2$ symmetry remains unbroken.  The lightest member of the $\phi_2$ doublet will then be stable.  We shall identify one of the neutral
members of $\phi_2$ as the WDM $\sigma$ with a mass of order keV.

Neutrino masses are generated via the seesaw mechanism.  Three $Z_2$ even singlet neutrinos, $N_i$, are introduced.
The Yukawa Lagrangian of the model is
\begin{equation}
{\cal L}_{\rm Yuk} = {\cal L}_{\rm Yuk}^{\rm SM} + (Y_N)_{ij} \ell_i N_j \,\phi_1 + \frac{M_{N_i}}{2}N_i^T C N_i + h.c.
\end{equation}
Here ${\cal L}_{\rm Yuk}^{\rm SM}$ is the SM Yukawa coupling Lagrangian and involves only the $\phi_1$ field owing to the
$\phi_2 \rightarrow -\phi_2$ reflection ($Z_2$) symmetry.  The Higgs potential of the model is
\begin{eqnarray}
V &=& -m_1^2 |\phi_1|^2 + m_2^2 |\phi_2|^2 + \lambda_1 |\phi_1|^4 + \lambda_2 |\phi_2|^4 + \lambda_3 |\phi_1|^2 |\phi_2|^2
\nonumber \\
&+& \lambda_4 |\phi_1^\dagger \phi_2|^2 + \{\frac{\lambda_5}{2}(\phi_1^\dagger \phi_2)^2 + h.c.\}.
\end{eqnarray}
With $\langle\phi_1^0\rangle = v \simeq 174$ GeV and $\langle \phi_2^0 \rangle = 0$, the masses of the various fields
are obtained as
\begin{eqnarray}
m_h^2 &=& 4 \lambda_1 v^2,~~~ m_\sigma^2 = m_2^2 + (\lambda_3+ \lambda_4 +\lambda_5)\, v^2; \nonumber \\
m_A^2 &=& m_2^2 + (\lambda_3 + \lambda_4 - \lambda_5)\, v^2;~~~ m_{H^\pm}^2 = m_2^2 + \lambda_3 v^2~.
\label{masses}
\end{eqnarray}
Here $h$ is the SM Higgs boson with a mass of 126 GeV;  $\sigma$ and $A$ are the second scalar and
pseudoscalar fields, while $H^\pm$ are the charged scalars.  We wish to identify $\sigma$ as the keV warm dark matter
candidate.\footnote{Alternatively, $A$ can be identified as the WDM candidate. With some redefinitions of couplings, this
scenario would lead to identical phenomenology as in the case of $\sigma$ WDM.}  An immediate concern is whether the other
scalars can all be made heavy, of order 100 GeV or above, to be consistent with experimental data.  This can indeed be done,
as can be seen from Eq. (\ref{masses}). Note that $m_A^2 = m_\sigma^2 - 2 \lambda_5 v^2$ and $m_{H^\pm}^2 = m_\sigma^2
 - (\lambda_4+\lambda_5)v^2$, so that even for $m_\sigma \sim$ keV, $m_A$ and $m_{H^\pm}$ can be large.
 However, the masses of $A$ and $H^\pm$ cannot be taken to arbitrary large
values, since  $\lambda_i v^2$ are at most of order a few hundred (GeV)$^2$ for perturbative values of $\lambda_i$.
The boundedness conditions on the Higgs potential can all be satisfied \cite{review} with the choice of positive $\lambda_{1,2,3}$ and negative $\lambda_5$ and $(\lambda_{4}+\lambda_5)$.
The keV WDM version of the two Higgs doublet model would thus predict that the neutral scalar $A$ and the charged scalar $H^\pm$ have masses not more than a few hundred GeV.  The present limits on the masses of $A$ and $H^\pm$ are approximately
$m_A > 90$ GeV (from $Z$ boson decays into $\sigma + A$) and $m_{H^\pm} > 100$ GeV from LEP searches for charged scalars.

\subsection{Electroweak precision data and Higgs decay constraints}

The precision electroweak parameter $T$ receives an additional contribution from the second Higgs doublet, which is given by\cite{inert}
\begin{equation}
\Delta T = \frac{m^2_{H^\pm}}{32 \pi^2 \alpha v^2} \left[1 - \frac{m_A^2}{m^2_{H^\pm}-m_A^2} {\rm log}\frac{m^2_{H^\pm}}{m_A^2}
\right]
\end{equation}
where the mass of $\sigma$ has been neglected.  For $\{m_{H^\pm}, \,m_A\} = \{150, \, 200\}$ GeV, $\Delta T \simeq -0.095$
while for $\{m_{H^\pm}, \,m_A\} = \{200, \, 150\}$ GeV, $\Delta T \simeq +0.139$. Both these numbers are consistent with current precision electroweak data constraints, $T = 0.01 \pm 0.12$ \cite{pdg}.  Note, however, that the mass splitting between $H^\pm$ and $A$ cannot be too much, or else the limits on $T$ will be violated. For example, if $\{m_{H^\pm}, \,m_A\} = \{150, \, 300\}$ GeV,
$\Delta T \simeq -0.255$, which may be disfavored.

The parameter $S$ receives a new contribution from the second Higgs doublet, which is evaluated to be
\begin{equation}
\Delta S = \frac{1}{12 \pi}\left({\rm log} \frac{m_A^2} {\,m^2_{H^\pm}} - \frac{5}{6} \right)~.
\end{equation}
If $\{m_{H^\pm}, \,m_A\} = \{150, \, 200\}$ GeV, $\Delta S \simeq +0.025$, while for  $\{m_{H^\pm}, \,m_A\} = \{200, \, 150\}$ GeV, $\Delta S \simeq -0.007$.  These values are consistent with precision electroweak data which has $S = -0.03 \pm 0.10$
\cite{pdg}.

In this model the decay $h \rightarrow \sigma \sigma$ can occur proportional to the quartic coupling combination
$(\lambda_3 + \lambda_4 + \lambda_5)$.  The decay rate is given by
\begin{equation}
\Gamma (h \rightarrow \sigma \sigma) = \frac{|\lambda_3+\lambda_4 + \lambda_5|^2}{16 \pi} \frac{v^2}{m_h}~.
\end{equation}
Since the invisible decay of the SM Higgs should have a branching ratio less than 23\% \cite{gunion}, we obtain the limit
(using $\Gamma = 4.2 \pm 0.08$ MeV for the SM Higgs width)
\begin{equation}
|\lambda_3 + \lambda_4+\lambda_5| < 1.4 \times 10^{-2}~.
\end{equation}

We thus see broad agreement with all experimental constraints in the two Higgs doublet models with a keV neutral scalar
identified as warm dark matter.

\subsection{Late decay of right-handed neutrino \boldmath{$N$}}

\label{latedecay}

Before proceeding to discuss the early universe cosmology within the two Higgs doublet model with warm dark matter, let us identify the parameter space of the model where the late decay of a particle occurs with a lifetime in the range of $(10^{-4}-1)$ sec. Such a decay is necessary in order to dilute the warm dark matter abundance within the model, which would otherwise be too large. A natural candidate for such late decays is one of the heavy right-handed neutrinos, $N$, that participates in the seesaw mechanism for small neutrino mass generation.  If its lifetime were longer than 1 sec. that
would affect adversely the highly successful big bang nucleosynthesis scheme. Lifetime shorter than $10^{-4}$ sec. would not
lead to efficient reheating of radiation in the present model, as that would also reheat the warm dark matter field.

It turns out that the masses and couplings of the late--decaying field $N$ are such that its contribution to the
light neutrino mass is negligibly small.  The smallest neutrino mass being essentially zero can be taken as one of the
predictions of the present model.  We can therefore focus on the mixing of this nearly decoupled $N$ field with light
neutrinos.  For simplicity we shall assume mixing of $N$ with one flavor of light neutrino, denoted simply as $\nu$.
The mass matrix of the $\nu-N$ system is then given by
\begin{eqnarray}
M_\nu = \left(\begin{matrix} 0 & Y_N v \\ Y_N v & M_N\end{matrix}    \right)~.
\label{seesaw}
\end{eqnarray}
A light--heavy neutrino mixing angle can be defined from Eq. (\ref{seesaw}):
\begin{equation}
\sin\theta_{\nu N} \simeq \frac{Y v}{M_N}~.
\label{angle}
\end{equation}
This mixing angle will determine the lifetime of $N$.

If kinematically allowed, $N$ would decay into $ h \nu,\, h \overline{\nu},\, W^+ e^-,\, W^- e^+,\, Z\nu$ and
$Z \overline{\nu}$.  These decays arise through the $\nu-N$ mixing.  The total two body decay rate of $N$ is given by
\begin{equation}
\Gamma(N \rightarrow h \nu,\, h \overline{\nu},\, W^+ e^-,\, W^- e^+,\, Z\nu,\;Z \overline{\nu}) = ~~~~~~~~~~~~~~~~~~~~~~~~~~~~~~~~~~~~~~~~~~~~`
\nonumber
\end{equation}
\begin{equation}
\frac{Y_N^2 M_N} {32 \pi} \left[\left(1-\frac{m_h^2}{M_N^2}\right)^2 + 2\left(1-\frac{m_W^2}{M_N^2}\right)^2\left(1+  \frac{2 m_W^2}{M_N^2}\right)
+ \left(1-\frac{m_Z^2}{M_N^2}\right)^2 \left(1+ \frac{2m_Z^2}{M_N^2}\right) \right] ~.
\label{2body}
\end{equation}

\noindent Here the first term inside the square bracket arises from the decays $N \rightarrow h \nu$ and $N \rightarrow h \overline{\nu}$, the second term from decays of $N$ into $W^\pm e^\mp$ and the last term from $N$ decays into
$Z \nu$ and $Z \overline{\nu}$.  We have made use of the expression for the mixing angle given in Eq. (\ref{angle}),
which is assumed to be small.

When the mass of $N$ is smaller than 80 GeV, these two body decays are kinematically not allowed.  In this case,
three body decays involving virtual $W$ and $Z$ will be dominant.  The total decay rate for $N$ decaying into three body final
states through the exchange of the $W$ boson is given by
\begin{equation}
\Gamma(N \rightarrow 3~{\rm body}) = \frac{G_F^2 M_N^5}{192 \pi^3} \sin^2\theta_{\nu N}\left(1 + \frac{3}{5} \frac{M_N^2}{m_W^2}\right) (2) \left[5 + 3 F\left(\frac{m_c^2}{M_N^2}\right) + F\left(\frac{m_\tau^2}{M_N^2}\right) \right]~.
\label{3body}
\end{equation}
This expression is analogous to the standard muon decay rate.  An overall factor of 2 appears here since $N$ being Majorana
fermion decays into conjugate channels. The factor 5 inside the square bracket accounts for the virtual $W^+$ boson decaying
into $e^+ \nu_e,\, \mu^+ \nu_\mu$ and $\overline{d} u$ for which the kinematic function $F(x) = \{1- 8 x + 8 x^3 - x^4-12 x^2\, {\rm ln} x\}$ is close to one \cite{kuno}.  For $M_N > 175$ GeV, an additional piece, $3F(m_t^2/M_N^2)$, should
be included inside the square bracket of Eq. (\ref{3body}). Analogous expressions for three body decay of $N$ via virtual $Z$ boson are found
to be numerically less important (about 10\% of the virtual $W$ contributions) and we ignore them here.  Virtual Higgs boson
exchange for three body $N$ decays are negligible owing to small Yukawa coupling suppressions.  We shall utilize expressions
(\ref{2body}) and (\ref{3body}) in the next section where the relic density of $\sigma$ WDM is computed.

\section{Relic Abundance of Warm Dark Matter \boldmath{$\sigma$} }

Here we present a calculation of the relic abundance of $\sigma$ which is taken to have a mass of order keV, and  which serves
as warm dark matter of the universe.  Since $\sigma$ has thermal abundance, it turns out that relic abundance
today is too large compared to observations.  This situation is remedied in the model by the late decay of
$N$, the right--handed neutrino present in the seesaw sector.  To see consistency of such a scheme, we should
follow carefully the thermal history of the WDM particle $\sigma$.

When the universe was hot, at temperatures above the $W$ boson mass, $\sigma$ was in thermal equilibrium via
its weak interactions through scattering processes such as $W^+ W^- \rightarrow \sigma \sigma$. As temperature
dropped below the $W$ boson mass, such processes became rare, since the number density of $W$ boson got depleted.
The cross section for the process $W^+ \sigma \rightarrow W^+ \sigma$ at energies below the $W$ mass is given by
\begin{equation}
\sigma(W^+ \sigma \rightarrow W^+ \sigma) \simeq \left(\frac{g^4}{64 \pi}\right)\frac{1}{m_W^2}~.
\end{equation}
The interaction rate $\langle \sigma n v\rangle$ is then given by
\begin{equation}
\langle n \sigma v \rangle \approx \left(\frac{g^4}{64 \pi }\right)\frac{1}{m_W^2} T^3 \left(\frac{m_W}{T}\right)^{3/2} e^{-m_W/T}
\end{equation}
where the Boltzmann suppression factor in number density of $W$ appears explicitly.  Demanding this interaction rate
to be below the Hubble expansion rate at temperature $T$, given by $H(T) = 1.66 g_*^{1/2} T^2/M_P$, with
$g_*$ being the effective degrees of freedom at $T$ and $M_P = 1.19 \times 10^{19}$ GeV, we obtain the
freeze--out temperature for this process to be $T_f \simeq 2.5$ GeV (with $g_* \approx 80$ used).

$\sigma$ may remain in thermal equilibrium through other processes.  The scattering $b \sigma \rightarrow b \sigma$
mediated by the Higgs boson $h$ of mass 126 GeV is worth considering.  ($b$ quark has the largest Yukawa coupling among
light fermions.)  The cross section for this process at energies below the $b$-quark mass is given by
\begin{equation}
\sigma (b \sigma \rightarrow b \sigma) \simeq \frac{|\lambda_3+\lambda_4+\lambda_5|^2 m_b^2}{4 \pi m_h^4}~.
\end{equation}
If $|\lambda_3+\lambda_4+\lambda_5| = 10^{-2}$, this process will freeze out at $T_f \approx 240$ MeV ($g_* = 70$ is
used in this estimate, along with Boltzmann suppression.)  For smaller values of $|\lambda_3+\lambda_4+\lambda_5|$, the freeze--out temperature will be higher.

The process $\mu^+ \mu^- \rightarrow \sigma \sigma$ mediated by the Higgs boson $h$ can keep $\sigma$ in thermal equilibrium down to lower temperatures,
since the $\mu^\pm$ abundance is not Boltzmann suppressed.  The cross section is given by
\begin{equation}
\sigma(\mu^+ \mu^- \rightarrow \sigma \sigma) = \frac{|\lambda_3+\lambda_4+\lambda_5|^2}{64\pi} \frac{m_\mu^2}{m_h^4}~.
\end{equation}
The number density of $\mu^\pm$, which are in equilibrium, is given by $0.2 T^3$, from which we find that this process
would go out of thermal equilibrium at $T \approx 250$ MeV for $|\lambda_3+\lambda_4+\lambda_5|= 10^{-2}$.  This process
could freeze out at higher temperatures for smaller values of $|\lambda_3+\lambda_4+\lambda_5|$.

\begin{center}
\begin{figure}
\hspace*{0.8in}
\subfloat{\includegraphics[width = 2.5in]{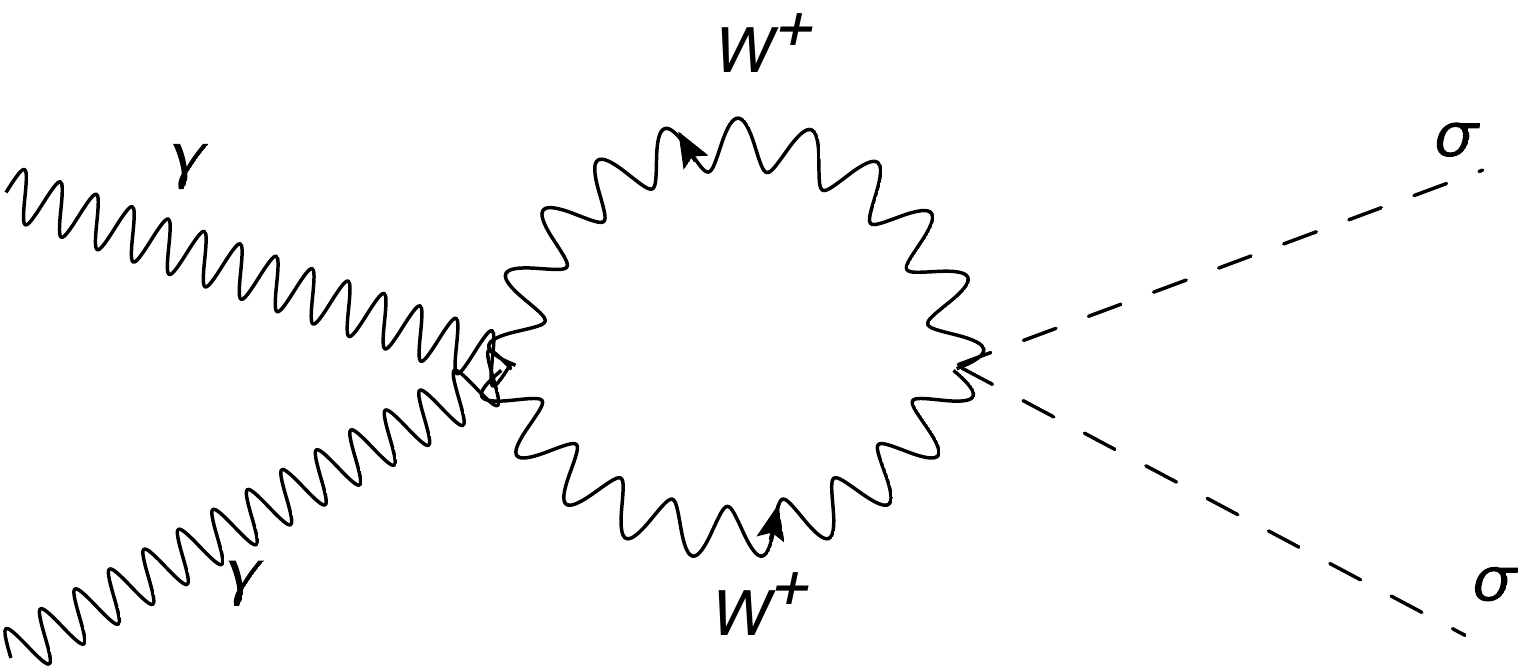}} \hspace*{0.1in}
\subfloat{\includegraphics[width = 2.5in]{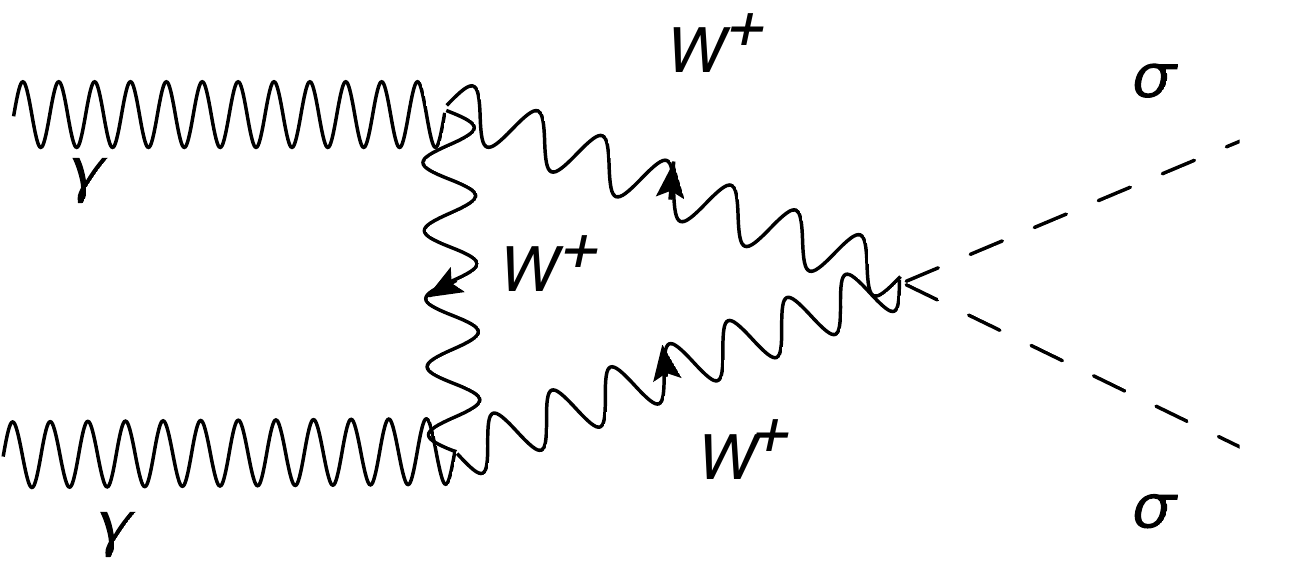}}
\vspace*{0.05in}
\caption{Loop diagrams leading to $\gamma \gamma \rightarrow \sigma \sigma$.}
\label{loop}
\end{figure}
\end{center}

There is one process which remains in thermal equilibrium independent of the values of the Higgs quartic couplings.
This is the scattering $\gamma \gamma \rightarrow \sigma \sigma$ mediated by the $W^\pm$ gauge bosons shown in Fig. \ref{loop}.  The relevant couplings are all fixed, so that the cross section has no free parameters.  We find it to be
\begin{equation}
\sigma(\gamma \gamma \rightarrow \sigma \sigma) = \frac{E_\sigma^2 F_W^2}{64 \pi} \left[\frac{e^2 g^2}{32 \pi^2 m_W^2}\right]^2
\label{prod}
\end{equation}
where $F_W = 7$ is a loop function. Using $E_\sigma = 3.15 T$ and $n_\gamma = 0.2 T^3$, the interaction rate $\langle
\sigma n v\rangle$ can be computed.  Setting this rate to be equal to the Hubble expansion rate, we find that this process
freezes out at $T \approx 150$ MeV (with $g_* = 17.25$ appropriate for this temperature used).  Among all scattering
processes, this one keeps $\sigma$ to the lowest temperature, and thus the freeze-out of $\sigma$ occurs at
$T_{f, \sigma} \approx 150$ MeV with a corresponding $g_*^\sigma = 17.25$.

Having determined the freeze--out temperature of $\sigma$ to be $T_f^\sigma \approx 150$ MeV, we can now proceed to
compute the relic abundance of $\sigma$.  We define the abundance of $\sigma$ as
\begin{equation}
Y_\sigma = \frac{n_\sigma}{s}
\end{equation}
where $n_\sigma$ is the number density of $\sigma$ and $s$ is the entropy density.  These two quantities are given
for relativistic species to be
\begin{equation}
n_\sigma = \frac{g_\sigma \zeta(3)}{\pi^2} T^3,~~~s = \frac{2 \pi^4}{45}g_{\rm efff} T^3,
\end{equation}
where
\begin{equation}
g_{\rm efff} = \sum_{\rm bosons}g_b + \frac{7}{8} \sum_{\rm fermions} g_f~.
\end{equation}
Thus
\begin{equation}
Y_\sigma = \frac{45 \zeta(3)}{2\pi^4} \frac{g_\sigma}{g_{\rm eff}}~.
\end{equation}
Since $Y_\sigma$ is a thermally conserved quantity as the universe cools, we can obtain the abundance of $\sigma$ today as
\begin{equation}
\Omega_\sigma = Y_\sigma m_\sigma \frac{s_0}{\rho_c},
\end{equation}
where $s_0 = 2889.2/{\rm cm}^3$ is the present entropy density and $\rho_c = 1.05368 \times 10^{-5} h^2 \,{\rm GeV}/{\rm cm}^3$
is the critical density.  Using $g_\sigma = 1$ appropriate for a real scalar field and with $h = 0.7$ we thus obtain
\begin{equation}
\Omega_\sigma = 9.02 \left(\frac{17.25}{\rm g_{\rm eff}}\right) \left(\frac{m_\sigma}{1 \, {\rm keV}}\right)~.
\label{abund0}
\end{equation}
Here we have normalized $g_{\rm eff} = 17.25$, appropriate for the freeze--out temperature of $\sigma$.  We see
from Eq. (\ref{abund0}) that for a keV warm dark matter, $\Omega_\sigma$ is a factor of 34 larger than the observed
value of $0.265$.  For a clear discussion of the relic abundance in a different context see Ref. \cite{zhang}.

\subsection{Dilution of \boldmath{$\sigma$} abundance via late decay of \boldmath{$N$}}

The decay of $N$ involved in the seesaw mechanism, as discussed in Sec. \ref{latedecay}, can dilute the abundance
of $\sigma$ and make the scenario consistent.  We assume that at very high temperature $N$ was in thermal equilibrium.
This could happen in a variety of ways.\footnote{Late decays of heavy particles have been used in order to dilute
dark matter abundance in other contexts \cite{lindner,zhang}.}  For example, one could have inflaton field $S$ couple to $N$ via a Yukawa coupling of type $SNN$
which would then produce enough $N$'s in the process of reheating after inflation~\cite{shafi}. Alternatively, the two Higgs extension of SM model could be an effective low energy theory which at high energies could have a local $B-L$ symmetry.
The $B-L$ gauge interactions would keep $N$ in thermal equilibrium down to temperatures a few times
below the gauge boson mass, at which point $N$ freezes out.  As the universe cools, the Hubble expansion rate also slows down.
The two body and three body decays of $N$, given in Eqs. (\ref{2body})-(\ref{3body}), will come into equilibrium at some
temperature at which time $N$ would begin to decay.  If this temperature $T_d$ is in the range of 150 MeV to 1 MeV, the decay
products (electron, muon, neutrinos, up quark and down quark) will gain entropy as do the photons which are in thermal
equilibrium with these species.\footnote{At $T= 150$ MeV, it is not completely clear if we
should include the light quark degrees of freedom or the hadronic degrees.  We have kept the $u$ and $d$ quarks in our
decay rate evaluations.}  Since $\sigma$ froze out at $T \approx 150$ MeV, and since $\sigma$ is not a decay product
of $N$, the decay of $N$ will cause the temperature of photons to increase relative to that of $\sigma$.  Thus a dilution
in the abundance of $\sigma$ is realized.  Note that the decay temperature $T_d$ should be above one MeV, so that big bang
nucleosynthesis is not affected.  The desired range for the lifetime of $N$ is thus $\tau_N = (10^{-4} - 1)$ sec.

The reheat temperature $T_r$ of the thermal plasma due to the decays of $N$ is given by \cite{turner}
\begin{equation}
T_r = 0.78 [g_*(T_r)]^{-1/4} \sqrt{\Gamma_N M_P}~.
\end{equation}
Energy conservation then implies the relation
\begin{equation}
M_N Y_N s_{\rm before} = \frac{3}{4} s_{\rm after} T_r~.
\end{equation}
If the final state particles are relativistic, as in our case, a dilution factor defined as
\begin{equation}
d=\frac{s_{\rm before}}{s_{\rm after}}
\end{equation}
takes the form
\begin{equation}
d = 0.58\, [g_*(T_r)]^{-1/4} \sqrt{\Gamma_N M_P}/(M_N Y_N)~.
\end{equation}
The abundance of $N$ is given by
\begin{equation}
Y_N = \frac{135}{4\pi^4} \frac{\zeta(3)}{g(T_{f,N})},
\end{equation}
where $g(T_{f,N})$ stands for the degrees of freedom at $N$ freeze--out.  Putting all these together we
obtain the final abundance of $\sigma$ as
\begin{equation}
\Omega_\sigma = (0.265) \left(\frac{m_\sigma}{1 \, {\rm keV}}\right) \left(\frac{7.87 \, {\rm GeV}}{M_N} \right)
\left(\frac{1\, {\rm sec.}}{\tau_N} \right)^{1/2} \left(\frac{g(T_{f,N})}{106.75}\right)\left(\frac{17.25}{g_f^\sigma} \right).
\label{abundance}
\end{equation}
Here we have normalized various parameters to their likely central values and used $g_*(T_r) = 10.75$.  The value of
$g(T_{f,N}) = 106.75$ counts all SM degrees and nothing else.

\begin{center}
\begin{figure}
\includegraphics[width = 6in]{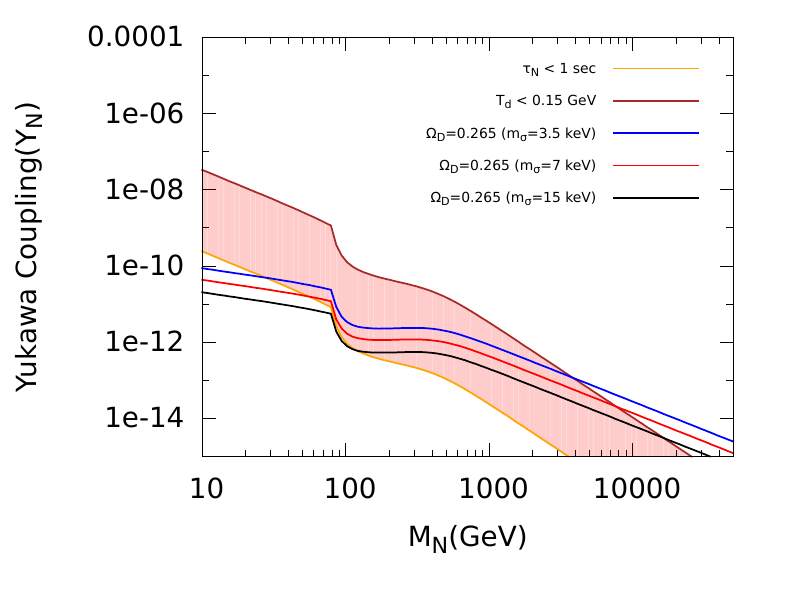}
\caption{Allowed parameter space of the model in the $M_N-Y_N$ plane. The shaded region corresponds to
the decay temperature $T_d$ of $N$ lying in the range 150 MeV -- 1 MeV.  The three solid curves generate
the correct dark matter density $\Omega_D$ for three different values of the WDM mass $m_\sigma = \{3.5,\, 7,\, 15\}$ keV.}
\label{plot}
\end{figure}
\end{center}

From Eq. (\ref{abundance}) we see that the correct relic abundance of $\sigma$ can be obtained for $M_N \sim 10$ GeV
and $\tau_N \sim 1$ sec.  In Fig. \ref{plot} we have plotted the dark matter abundance as a function of $M_N$ and its Yukawa coupling $Y_N$
for three different values of $m_\sigma$ (3.5, 7 and 15 keV).  Also shown in the figure are the allowed band for
$\tau_N$ to lie in the range of $(10^{-4} - 1)$ sec., or equivalently for $T_d = (150-1)$ MeV.  We see that there is a significant region allowed by the model
parameters.  We also note that the mass of $N$ should lie in the range $M_N = 25 \,{\rm GeV} - 20 \, {\rm TeV}$
for the correct abundance of dark matter.

A remark on the average momentum $\langle p_\sigma\rangle$ of the dark matter is in order.  The dilution factor
$d \simeq 1/34$ for $m_\sigma = 1$ keV.  The temperature of $\sigma$ is thus cooler by a factor of $1/(34)^{1/3} =
0.31$ relative to the photon.  The momentum of $\sigma$ gets redshifted by a factor $\xi^{-1/3} = 0.58$ where
$\xi = g_f^\sigma/g_{\rm today} = 17.25/3.36$.  The net effect is to make $\langle p_\sigma \rangle/(3.15 T) = 0.18$.

\section{Other Implications of the Model}

In this section we discuss briefly some of the other implications of the model.

\subsection{Supernova energy loss}

The process $\gamma \gamma \rightarrow \sigma \sigma$ can lead to the production of $\sigma$ inside supernova core.
Once produced these particles will freely escape, thus contributing to new channels of supernova energy loss.
Note that $\sigma$ does not have interactions with the light fermions. The cross section for $\sigma$ production is
given in Eq. (\ref{prod}).  Here we make a rough estimate of the energy lost via this process and ensure that this
is not the dominant cooling mechanism of supernovae. We follow the steps of Ref. \cite{bm} here.
The rate of energy loss is given by
\begin{equation}
Q = V_{\rm core} n_\gamma^2 \langle E \rangle \sigma(\gamma \gamma \rightarrow \sigma \sigma)
\end{equation}
where $V_{\rm core} = 4\pi R_{\rm core}^3/3$ is the core volume and we take $R_{\rm core} = 10$ km.
$n_\gamma \simeq 0.2 T_\gamma^3$ is the photon number density, and $\langle E \rangle = 3.15 T_\gamma$ is the
average energy of the photon.  Using Eq. (\ref{prod}) for the cross section we obtain $Q \sim 2.8 \times 10^{51}$ erg/sec,
when $T_\gamma = 30$ MeV is used.  Since the supernova explosion from 1987A lasted for about 10 seconds, the total
energy loss in $\sigma$ would be about $2.8 \times 10^{52}$ erg, which is to be compared with the total energy loss
of about $10^{53}$ erg.  This crude estimate suggests that energy loss in the new channel is not excessive.
We should note that the energy loss scales as the ninth power of core temperature, so for larger values of
$T_\gamma$, this process could be significant. A more detailed study of this problem would be desirable.

\subsection{Dark matter self interaction}

In our model dark matter self interaction, $\sigma \sigma \rightarrow \sigma \sigma$, occurs proportional to
$|\lambda_2|^2$.  There are
rather severe constraints on self-interaction of dark matter from dense cores of galaxies and
galaxy clusters where the velocity distribution can be isotropized. Constraints from such
halo shapes, as well as from dynamics of bullet cluster merger have been used to infer an
upper limit on the dark matter self-interaction cross section \cite{manoj}:
\begin{equation}
\frac{\sigma}{m_\sigma} < 1~ {\rm barn}/{\rm GeV}~.
\end{equation}
The self interaction cross section in the model is given by
\begin{equation}
\sigma (\sigma \sigma \rightarrow \sigma \sigma) = \frac{9 \hat{\lambda}_2^2}{8 \pi s}~
\end{equation}
where $\hat{\lambda_2} = \lambda_2 - |\lambda_3+\lambda_4+\lambda_5|^2 \,(v^2/m_h^2)$, with the
second term arising from integrating out the SM Higgs field $h$.
This leads to a limit on the coupling $\hat{\lambda}_2$ given by
\begin{equation}
\hat{\lambda}_2 < 5.4 \times 10^{-6} \left(\frac{m_\sigma}{10\,{\rm keV}} \right)^{3/2}~.
\end{equation}
The one loop corrections to $\sigma$ self interaction strength is of order $g^4/(16 \pi^2)\sim 10^{-3}$. So we use the tree level $\lambda_2$ to cancel this to make the effective self interaction strength of order $10^{-6}$ as needed.
One can choose $|\lambda_3+\lambda_4+\lambda_5| \sim 10^{-3}$, so that the effective quartic coupling $\hat{\lambda_2}$
is positive.

\subsection{The extra-galactic $X$-ray anomaly}

Recently two independent groups have reported the observation of a peak in the extra-galactic
$X$-ray spectrum at 3.55 keV \cite{Xray1,Xray2}, which appear to be not understood in terms of known physics and
astrophysics. While these claims still have to be confirmed by other observations, it is tempting to speculate
that they arise from the decay of WDM into two photons.  If the $Z_2$ symmetry remains unbroken, $\sigma$ is
absolutely stable in our model and will not explain this anomaly.  However, extremely tiny breaking of this symmetry via a soft term of the type
$\phi_1^\dagger \phi_2$ can generate the reported signal.  Such a soft breaking term would induce a nonzero
vacuum expectation value for $\sigma$ which we denote as $u$.
In order to explain the $X$-ray anomaly, this VEV has to be in the range
$u = (0.03-0.09)$ eV.  This comes about from the decay rate, which is given by
\begin{equation}
\Gamma(\sigma \rightarrow \gamma \gamma) = \left(\frac{\alpha}{4\pi} \right)^2 F_W^2 \left(\frac{u^2}{v^2} \right)
\frac{G_F m_\sigma^3}{8 \sqrt{2} \pi}
\end{equation}
with $F_W = 7$, which is matched to a partial lifetime in the range $\Gamma^{-1} (\sigma \rightarrow \gamma \gamma) = (4 \times 10^{27} -
4 \times 10^{28})$ sec \cite{Xray1,Xray2}.  Once $\sigma$ develops a vacuum expectation value, it also mixes with
SM Higgs field $h$, but this effect is subleading for the decay $\sigma \rightarrow \gamma \gamma$.  Such mixing
was the main source of the two photon decay of WDM in the case of a singlet scalar WDM of Ref. \cite{bm}.

\subsection{Collider signals}

The charged scalar $H^\pm$ of the model can be pair produced at the Large Hadron Collider via the Drell-Yan process.
$H^+$ will decay into a  $W^+$ and a $\sigma$. This signal has been analyzed in Ref. \cite{rai} within the context
of a similar model \cite{nandi}.  Sensitivity for these charged scalars would require 300 $fb^{-1}$ of luminosity
of LHC running at 14 TeV.

The pseudoscalar $A$ can be produced in pair with a $\sigma$ via $Z$ boson exchange. $A$ will decay into
a $\sigma$ and a $Z$.  The $Z$ can be tagged by its leptonic decay.  Thus the final states will have two leptons
and missing energy.  The Standard Model $ZZ$ background with the same final states would be much larger.
We can make use of the fact that in the signal events, the $Z$ boson which originates from the decay $A \rightarrow Z \sigma$
with a heavy $A$ and a massless $\sigma$ will be boosted in comparison with the background $Z$ events.
This  will  reflect in the $p_T$ distribution which would be
different for the signal events compared to the SM background $Z$'s.
Studies to look for this kind of signals in this particular framework are in order.

\section{Conclusions}
In this paper we have proposed a novel warm dark matter candidate in the context of two Higgs doublet extensions of the
Standard Model.  We have shown that a neutral scalar boson of these models can have a mass in the keV range.  The abundance of
such a thermal dark matter is generally much higher than observations; we have proposed a way to dilute this by the late
decay of a heavy right-handed neutrino which takes part in the seesaw mechanism.  A consistent picture emerges where
the mass of $N$ is in the range 25 GeV to 20 TeV.  The model has several testable consequences at colliders as well
as in astrophysical settings.  The charged scalar and the pseudoscalar in the model cannot be much heavier than a few
hundred GeV.  It will be difficult to  see such a warm dark matter candidate in direct detection experiments.
Supernova dynamics may be significantly modified by the production of $\sigma$ pairs in photon--photon
collisions.  The model can also explain the anomalous $X$-ray signal reported by different groups in the extra-galactic
spectrum.

\section*{Acknowledgments}
We thank Ernest Ma and Xerxes Tata for helpful discussions.
This research is supported in part by the National Science Foundation under Grant No. NSF PHY11-25915 (KSB and SS).
The work of KSB is supported in part by the US Department of Energy Grant No.
de-sc0010108 and RNM is supported in part by the National Science Foundation Grant No.
PHY-1315155. SS is supported by a KITP Graduate Fellowship.


\begin{thebibliography}{99}

\bibitem{susy}
For a review see for e.g:
 S.~P.~Martin, ``A Supersymmetry primer,''
  Adv.\ Ser.\ Direct.\ High Energy Phys.\  {\bf 21}, 1 (2010)
  [hep-ph/9709356].

  \bibitem{lr}
 R.~N.~Mohapatra and G. Senjanovi\'c
 Phys.\ Rev.\ Lett.\ {\bf 44}, 912 (1980).

  \bibitem{dfsz}
   M.~Dine, W.~Fischler and M.~Srednicki,
  Phys.\ Lett.\ B {\bf 104}, 199 (1981);
  A.~R.~Zhitnitsky,
  Sov.\ J.\ Nucl.\ Phys.\  {\bf 31}, 260 (1980)
  [Yad.\ Fiz.\  {\bf 31}, 497 (1980)].

  \bibitem{lee}
   T.~D.~Lee,
  Phys.\ Rev.\ D {\bf 8}, 1226 (1973).

  \bibitem{review}
  For a recent review see:
   G.~C.~Branco, P.~M.~Ferreira, L.~Lavoura, M.~N.~Rebelo, M.~Sher and J.~P.~Silva,
  Phys.\ Rept.\  {\bf 516}, 1 (2012)
  [arXiv:1106.0034 [hep-ph]].

  \bibitem{ma}

  E.~Ma,
  Phys.\ Rev.\ D {\bf 73}, 077301 (2006)
  [hep-ph/0601225].

  \bibitem{inert}
   R.~Barbieri, L.~J.~Hall and V.~S.~Rychkov,
  Phys.\ Rev.\ D {\bf 74}, 015007 (2006)
  [hep-ph/0603188].

  \bibitem{CDM}
  For a review of the problems of CDM at small scales see for e.g:
   D.~H.~Weinberg, J.~S.~Bullock, F.~Governato, R.~K.~de Naray and A.~H.~G.~Peter,
  ``Cold dark matter: controversies on small scales,''
  arXiv:1306.0913 [astro-ph.CO].

  \bibitem{WDM} S. Dodelson and L.M. Widrow, Phys.\ Rev.\ Lett.\ {\bf  72}, 17  (1994);
	X.D. Shi and G.M. Fuller, Phys.\ Rev.\ Lett.\ {\bf  82}, 2832 (1999);
	K. Abazajian, G.M. Fuller and M. Patel, Phys.\ Rev.\  {\bf D64}, 023501 (2001).

\bibitem{kusenko}
See for e.g.,  A.~Kusenko,
  Phys.\ Rept.\  {\bf 481}, 1 (2009)
  [arXiv:0906.2968 [hep-ph]].

\bibitem{bm}
   K.~S.~Babu and R.~N.~Mohapatra,
  Phys.\ Rev.\ D {\bf 89}, 115011 (2014)
  [arXiv:1404.2220 [hep-ph]].


\bibitem{pdg}

 K.~A.~Olive {\it et al.}  [Particle Data Group Collaboration],
  Chin.\ Phys.\ C {\bf 38}, 090001 (2014).

  \bibitem{gunion}

  G.~Belanger, B.~Dumont, U.~Ellwanger, J.~F.~Gunion and S.~Kraml,
  Phys.\ Lett.\ B {\bf 723}, 340 (2013)
  [arXiv:1302.5694 [hep-ph]].

  \bibitem{kuno}

  See for e.g.,
   Y.~Kuno and Y.~Okada,
  Rev.\ Mod.\ Phys.\  {\bf 73}, 151 (2001)
  [hep-ph/9909265].



  \bibitem{zhang}

   M.~Nemevsek, G.~Senjanovic and Y.~Zhang,
  JCAP {\bf 1207}, 006 (2012)
  [arXiv:1205.0844 [hep-ph]].

  \bibitem{lindner}

  F.~Bezrukov, H.~Hettmansperger and M.~Lindner,
  Phys.\ Rev.\ D {\bf 81}, 085032 (2010)
  [arXiv:0912.4415 [hep-ph]].

  \bibitem{shafi} V.~N.~Senoguz and Q.~Shafi,
  Phys.\ Lett.\ B {\bf 582}, 6 (2004).

  \bibitem{turner}

   R.~J.~Scherrer and M.~S.~Turner,
  Phys.\ Rev.\ D {\bf 31}, 681 (1985).


  \bibitem{manoj}

   M.~Kaplinghat, R.~E.~Keeley, T.~Linden and H.~B.~Yu,
  Phys.\ Rev.\ Lett.\  {\bf 113}, 021302 (2014)
  [arXiv:1311.6524 [astro-ph.CO]];
  M.~Kaplinghat, S.~Tulin and H.~B.~Yu,
  arXiv:1308.0618 [hep-ph].

  \bibitem{Xray1}
   E.~Bulbul, M.~Markevitch, A.~Foster, R.~K.~Smith, M.~Loewenstein and S.~W.~Randall,
  Astrophys.\ J.\  {\bf 789}, 13 (2014)
  [arXiv:1402.2301 [astro-ph.CO]].

  \bibitem{Xray2}

  A.~Boyarsky, O.~Ruchayskiy, D.~Iakubovskyi and J.~Franse,
  Phys.\ Rev.\ Lett.\  {\bf 113}, 251301 (2014)
  [arXiv:1402.4119 [astro-ph.CO]].

  \bibitem{rai}

  U.~Maitra, B.~Mukhopadhyaya, S.~Nandi, S.~K.~Rai and A.~Shivaji,
  Phys.\ Rev.\ D {\bf 89}, 055024 (2014)
  [arXiv:1401.1775 [hep-ph]].

  \bibitem{nandi}

   S.~Gabriel and S.~Nandi,
  Phys.\ Lett.\ B {\bf 655}, 141 (2007)
  [hep-ph/0610253].


\end{thebibliography}
\end{document}